\documentclass[prb,twocolumn,floatfix,showpacs,showkeywords]{revtex4}

\usepackage{amssymb}
\usepackage{amsmath}
\usepackage{amsfonts}
\usepackage{bm}
\usepackage{graphicx}

\renewcommand{\S}{{\mathcal{S}}}

\newcommand{\C}{{\mathcal{C}}}
\newcommand{\Ham}{{\hat{\mathcal{H}}}}

\newcommand{\bhi}{{\hat{b}_i}}
\newcommand{\bhj}{{\hat{b}_j}}
\newcommand{\bhdi}{{\hat{b}^\dagger_i}}

\newcommand{\bk}{{\bm k}}
\newcommand{\bx}{{\bm x}}

\newcommand{\br}{{\bm r}}

\newcommand{\hc}{{\textrm{h.c.}}}
\newcommand{\ud}{{\textrm{d}}}

\newcommand{\nn}{{\nonumber}}

\begin{document}

\title{Two distinct Mott-Insulator to Bose-glass transitions and breakdown of self averaging in the disordered Bose-Hubbard model}
\author{Frank Kr\"uger$^{1,2}$}
\author{Seungmin Hong$^1$}
\author{Philip Phillips$^1$}
\affiliation{$^1$Department of Physics, University of Illinois, 1110 W. Green
St., Urbana, IL 61801, USA\\
$^2$School of Physics and Astronomy, University of St. Andrews, St. Andrews, Fife KY16 9SS, United Kingdom}

\date{\today}

\begin{abstract}
We  investigate the instabilities of the Mott-insulating phase of the weakly disordered Bose-Hubbard model within a renormalization group analysis of the 
replica field theory obtained by a strong-coupling expansion around the atomic limit. We identify a new order parameter and associated correlation length scale that is
capable of capturing the transition from a state with zero compressibility, the Mott insulator, to one in which the compressibility is finite, the Bose glass.   The order 
parameter is the relative variance of the disorder-induced mass distribution.  In the Mott insulator, the relative variance renormalizes to zero, whereas it diverges in the Bose glass. 
The divergence of the relative variance signals the breakdown of self-averaging.  The length scale governing the breakdown of self-averaging is the distance between rare regions.  
This length scale is finite in the Bose glass but diverges at the transition to the Mott insulator with an exponent of $\nu=1/D$ for incommensurate fillings. Likewise, the  
compressibility vanishes with an exponent of $\gamma=4/D-1$ at the transition. At commensurate fillings, the transition is controlled by a different fixed point 
at which both the disorder and interaction vertices are relevant. 
\end{abstract}

\pacs{05.30.Jp, 72.15.Rn, 64.70.Tg, 64.60.ae}
\maketitle

\section{Introduction}
\label{sec.intro}

Two distinct mechanisms localize interacting bosons moving in a random environment.  Either repulsions from local on-site interactions on a $D$-dimensional lattice lock the 
bosons in place or disorder inhibits tunneling from site to site. The former insulating state is an incompressible Mott insulator (MI), while the second is a compressible Bose glass (BG).  
In this paper, we identify an order parameter and an associated length scale that can describe the vanishing of the compressibility at the MI/BG
transition.  The order parameter is the relative variance of the disorder-induced mass distribution. The key insight leading to this observation is that despite the relevance of the disorder 
vertex in $D<4$ spatial dimensions, it is possible to identify a Mott-insulating fixed point where the mass, corresponding to the Mott gap, diverges fast enough such that the relative 
variance of the mass distribution is renormalized to zero. Sufficiently close to the MI, both the mean and the relative variance diverge, identifying the adjacent phase as an 
insulating BG, irrespective of the boson filling. We will show that the length scale underlying this order parameter is the effective distance between rare regions.

In their initial treatment of this problem, known as the disordered Bose-Hubbard (BH) model, Fisher and colleagues\cite{Fisher+89} situated the BG between the MI and superfluid (SF), thereby 
preventing a direct SF/MI phase transition. They  pointed out however, that it is in principle possible but extremely unlikely that for sufficiently weak disorder the BG phase is completely 
suppressed at commensurate boson fillings. For more than twenty years, this question remained controversial as simulations and analytical arguments both support\cite{Scalettar+91,Krauth+91,Singh+92,Pai+96,Kisker+97,Pazmandi+98,Sen+01,Lee+01,Wu+08,Bissbort+09} and negate\cite{Mukhopadhyay+96,Freericks+96,Svistunov96,Herbut97,Herbut98,Prokofev+04,
Weichman+08} the possibility of a direct MI/SF transition in the presence of weak disorder. Only recently, strong arguments\cite{Pollet+09} based on a mathematical theorem have 
been presented for the existence of the BG upon the destruction of the MI, thereby precluding a direct transition to the SF. Given this result, it is important to establish precisely how 
the MI to BG transition obtains.  In this paper, we are concerned  with the critical theory for the MI/BG transition for strongly interacting bosons on a D-dimensional lattice subject to weak 
disorder.  Of particular interest is the universality of the transition, e.g. the identification of a diverging length scale and the determination of the critical exponent describing the divergence 
of the inverse compressibility. Despite extensive numerical\cite{Scalettar+91,Krauth+91,Pai+96,Kisker+97,Sen+01,Lee+01,Prokofev+04} and analytical\cite{Singh+92,
Mukhopadhyay+96,Freericks+96,Svistunov96,Herbut97,Herbut98,Pazmandi+98,Weichman+08,Wu+08,Bissbort+09,Pollet+09,Kruger+09} studies of the disordered
BH model, the nature of the MI/BG transition remains elusive.

To be more specific, in the following we consider the disordered BH model in its simplest form,
\begin{eqnarray}
\Ham & = & -t\sum_{\langle i,j\rangle}(\bhdi\bhj+\hc)+\sum_i(\epsilon_i-\mu)\hat{n}_i\nn\\
& & +\frac{U}{2}\sum_i\hat{n}_i(\hat{n}_i-1),
\label{BHmodel}
\end{eqnarray}
describing bosons with creation and annihilation operators $\bhdi$, $\bhi$ ($\hat{n}_i=\bhdi\bhi$)  which are hopping with amplitude $t$ between 
neighboring sites $i,j$ on a $D$-dimensional hyper-cubic lattice and which are subject to an on-site repulsion $U$. The chemical potential $\mu$ controls the boson filling and 
disorder enters via random uncorrelated on-site potentials $\epsilon_i$ which must be appropriately bounded for the model to exhibit stable MI phases.\cite{Fisher+89}

Since the formation of the MI is a consequence of strong repulsive interactions, the MI/SF transition in the clean system can only be understood within an effective long-wavelength theory 
which is dual to the original BH model (\ref{BHmodel}) and derived by a strong coupling expansion around the atomic limit.\cite{Fisher+89,Sachdev99,Sengupta+05} 
The mass in the theory is given by the Mott gap which vanishes at the transition and turns negative in the SF. In the latter broken-symmetry state the finite bosonic order parameter 
is proportional to the SF density. Although the underlying strong coupling expansion breaks down in the SF where the renormalization group (RG) flow is towards weak coupling,\cite{Kruger+09}
the effective long-wavelength theory can be used to analyze the instability of the MI. Since the theory is controlled in the MI state 
it should serve as a starting point to also analyze the instability towards the formation of the BG in the presence of weak disorder. 

At the outset, it would seem that an effective theory in terms of the bosonic order parameter is completely inapplicable to the MI/BG transition since in both insulating phases
the mass gap is finite and the SF correlations are short ranged. Certainly, the relevant length scale at the MI/BG transition has nothing to do with the SF correlation length but instead 
should be intrinsically related to the disorder. Such a divergent length scale should exist since the transition is expected to be continuous with a divergent 
inverse compressibility. In this work, we demonstrate that the order parameter for the MI/BG transition can be extracted from the disorder- averaged replica version of 
the bosonic order parameter theory. The relevant quantity is the relative variance of the disorder-induced mass distribution which renormalizes to zero in the MI and diverges in the BG 
and hence serves as the order parameter for the MI/BG transition. The corresponding length scale is the distance between rare regions in the BG which causes 
the breakdown of self-averaging in the system as indicated by a diverging relative variance. This correlation length is finite in the BG and diverges on approaching 
the MI.  

From the analysis of the RG flow of the mean value and relative variance of the mass distribution we obtain the key results that (i) the instability of the MI is always  
towards a BG, in agreement with recent arguments,\cite{Pollet+09} and that (ii) the MI/BG
transitions at incommensurate and commensurate boson fillings are not in the same universality class. In the former case of incommensurate
fillings, we calculate the critical exponents at 1-loop order: $\nu=1/D$ for the correlation length and $\gamma=4/D-1$ for the inverse compressibility.

The paper is organized as follows. In Sec.~\ref{sec.selfaveraging}, we argue that the relative variance of the disorder-induced mass distribution serves as the order parameter for the MI/BG transition which vanishes in the MI and acquires a finite value in the BG. We further identify the corresponding correlation length as the distance between rare regions in the BG which cause 
a breakdown of self-averaging in the system. The disorder-averaged replica theory describing the long-wavelength physics at strong coupling is derived in Sec.~\ref{sec.replicatheory}.
In Sec.~\ref{sec.RG}, the RG equations are reviewed and reformulated in a new set of variables which permits us to extract the scale dependence of the relative variance of
the induced random-mass distribution. The results are presented in Sec.~\ref{sec.results}. The phase boundary between the MI and the BG is determined by a numerical integration 
of the RG equations. For incommensurate boson fillings, we obtain analytical results for the correlation length and compressibility exponents. Finally, we show that the 
MI/BG at commensurate boson fillings is in a different universality class. In Sec.~\ref{sec.final} our main results are discussed.

\section{Breakdown of Self-Averaging}
\label{sec.selfaveraging}

The MI in the disordered BH model exists in the strong coupling limit.  While the disorder must be appropriately
bounded for the MI to persist,\cite{Fisher+89} the key limit that defines strong coupling is the ratio $U/t\gg1$, where $U$ is the on-site energy and $t$ the
hopping matrix element.  As with all strong coupling problems, the natural variables that uncloak the physics are not related straightforwardly to those in the UV-complete
Hamiltonian. Further, if a critical theory is to correctly describe the destruction of the MI, it should contain the seeds of the BG phase, which Fisher and 
co-workers\cite{Fisher+89} argued are due to the physics of rare regions. Consequently, the underlying critical theory might have nothing to do with the bare bosonic propagator 
and the associated superfluid correlation length but rather a new length scale that is intrinsically related to the disorder.  

Consider the atomic limit of this problem.  A scaling analysis around this regime\cite{Pazmandi+97} demonstrates that the correlation length, $\xi$, defined as the length
scale beyond which the system encounters rare-region physics is finite in the BG and diverges at the transition to the MI with an exponent of $\nu=1/D$ for generic bounded disorder 
distributions such as the box distribution $P(\epsilon)=1/(2\Delta)$ for $|\epsilon|\le\Delta$ and $P(\epsilon)=0$ for $|\epsilon|>\Delta$.
The violation of the lower bound of $\nu\ge 2/D$ known as the quantum Harris criterium\cite{Chayes+86} was attributed to a breakdown of self-averaging\cite{Pazmandi+97} in the BG 
phase. That is, the fluctuations are not governed by the central-limit theorem.  Several years ago, Aharony and Harris\cite{Harris+96} showed
that a break-down of self-averaging implies that the relative variance of any thermodynamic quantity must be finite.  Taken in tandem, these results imply that characterizing a transition 
governed by the physics of rare regions requires two quantities to be finite: 1) a finite length scale, $\xi$, over which a rare region is encountered and 2) a finite value of
the relative variance of any thermodynamic quantity. 

The central idea we advance here is that the nature of the MI/BG transition can be understood using perturbative RG techniques but with a {\it new set} of
variables and length scales that contain the information about the rare regions. The mass $r$ in the effective long wavelength theory in the clean system corresponds to the Mott gap 
which renders the correlations of the bosonic order parameter short ranged. Any form of disorder in the microscopic BH model, e.g. the potential disorder $\epsilon_i$ following a distribution $P(\epsilon)$, will induce an effective disorder distribution $\tilde{P}(r)$ of the mass coefficient $r$. Since in any insulating phase the superfluid order 
parameter must vanish on large scales, in both phases, the MI and the BG, the mean $\bar r$ of the distribution must diverge under the RG, $\bar r(\ell)\rightarrow \infty$.
Hence, it cannot be used to distinguish between the MI and BG phases.  Given the general considerations above, the natural quantity to distinguish the BG from the MI is the relative 
variance $R_r$ of the mass distribution,
\begin{equation}
R_r=\frac{\overline{(r-\bar{r})^2}}{\bar{r}^2}.
\label{rel.var}
\end{equation}

In the MI phase, the system is self averaging and therefore $R_r(\ell)\rightarrow 0$ since relative variances of extensive quantities have to vanish in the thermodynamic limit as a consequence 
of the central-limit theorem. Since the BG is characterized by a breakdown of self averaging, $R_r$ is non-zero in the thermodynamic limit and therefore serves as the order parameter 
for the MI/BG transition. In this paper, break down of self-averaging refers entirely to the fact that $R_r\ne 0$.  Note, $R_r(\ell)\rightarrow 0$ in the MI implies that $\xi$ diverges at the transition; that is, there is no finite length scale over which rare regions exist.  
We point out that a break-down of self averaging, as defined here, has nothing to do with a finite-size effect.  Rather, it is a function entirely of whether or not in the \emph{thermodynamic limit},
$\xi$, the distance over which rare regions exist, is finite.  If it is, $R_r$ is necessarily non-zero as we will demonstrate by an explicit calculation. 

To work out the RG flow
of $R_r$, we will use that the disorder vertex in the disorder-averaged replica theory is proportional to the variance of the mass distribution. 
We will proceed by recasting the RG equations  in terms of $R_r$ and show that these equations admit a stable MI fixed point
at $\bar{r}=\infty$ and $R_r=0$ as well as a critical MI/BG fixed point at $\bar{r}=\infty$ and finite $R_r$. We point out that such an identification of fixed points is only possible because 
of the identification of the relevant variables. Whereas the standard equations that are used to study this problem only capture a runaway flow away from the MI/SF fixed point of the clean 
system,\cite{Fisher+89,Kruger+09} in the new variables we find that the unstable MI/SF fixed point is connected with the new critical MI/BG fixed point by a separatrix. Our analysis will further demonstrate that the MI/BG transition at commensurate fillings belongs to a different universality class.

\section{Effective Replica Theory}
\label{sec.replicatheory}

We analyze the instability of the MI state towards the formation of the BG using the disorder-averaged replica version of the effective long-wavelength theory 
which is derived by an expansion around the atomic limit of the stable MI state. In this theory, the replica mixing disorder vertex is proportional to the variance of the mass distribution 
which allows us to extract the RG flow of the relative variance $R_r$ (\ref{rel.var}) which we have identified as the order parameter for the MI/BG transition.
In the absence of disorder, the effective action capturing the long-wavelength physics at strong coupling, $t/U\ll 1$,  
\begin{equation}
\S = \int_{\bx,x_0}\left( \gamma_1\phi^*\partial_0\phi+\gamma_2|\partial_0\phi|^2+
|\nabla\phi|^2+r|\phi|^2+h|\phi|^4\right),
\end{equation}
can be constructed using a coherent state path-integral representation of the BH model (\ref{BHmodel}) and decoupling the off-diagonal hopping terms with a 
Hubbard-Stratonovich (HS) transformation. In this action, we have rescaled the length and imaginary time $\tau$ to dimensionless units, $\bx=\Lambda\br$ with $\Lambda$ the momentum 
cutoff and $x_0=U\tau$. The precise dependence of the coupling constants on the microscopic parameters is known 
analytically.\cite{Sengupta+05} For example, the mass coefficient which corresponds to the Mott gap, is given by 
\begin{equation}
r=1-y\left( \frac{m+1}{m-x}+\frac{m}{1-m+x}\right)
\end{equation} 
with $x=\mu/U$, $y=2Dt/U$, and $m$ the number of bosons per site. The interaction vertex $h$ is proportional to $y^2$ reflecting the underlying strong-coupling expansion 
around the atomic limit ($y=0$). An explicit expression for $h$ can be found elsewhere.\cite{Sengupta+05,Kruger+09} At the mean-field level, the MI/SF phase boundary is 
determined by $r=0$.  The MI states for different fillings $m$ are characterized by a finite Mott gap, $r>0$, and $\langle\phi\rangle=0$. The SF obtains for $r<0$ where the SF 
density is proportional to $\langle\phi\rangle\neq 0$.

As a consequence of the $U(1)$ symmetry of the fields,\cite{Sachdev99} the temporal gradient terms are given by $\gamma_1=-\partial r/\partial x$ and 
$\gamma_2=-\frac 12\partial^2 r/(\partial x)^2$ and are therefore related to the slope and the curvature of the mean-field phase boundary.  At the tips of the MI lobes at values $x_m$ of 
the chemical potential where the boson filling is commensurate with the lattice and the MI states are most stable, $\gamma_1=0$, and the critical theory is characterized by a 
dynamical exponent $z=1$. At incommensurate fillings, $\gamma_1\neq 0$, corresponding to $z=2$ dynamics.  

Consider now the disordered case. The site energies $\epsilon_i$ will be uncorrelated and chosen from a bounded distribution, $\epsilon_i\in[-\Delta,\Delta]$.   
The identical HS transformation can be performed\cite{Kruger+09} yielding a dual theory with coefficients given by the expressions for the clean system but with the chemical 
potential $x=\mu/U$ shifted by the random potential $\epsilon_i/U$ on each site, e.g. $r_i=r(x-\epsilon_i/U)$ for the mass term.  
This expression is well defined on every site if the disorder distribution $P(\epsilon_i)$ is bounded, and the chemical potential $x$ lies in the intervals $[m-1+\delta,m-\delta]$ 
and $\delta=\Delta/U<1/2$.  Note that this stability condition ensures that the system exhibits stable MI phases in the presence of disorder.  

The procedure for constructing the effective action is now standard:  restore translational symmetry by performing the disorder average of the free energy using the replica 
trick and and take the continuum limit. In the resultant expression,
\begin{equation}
\S_\textrm{eff} =  \sum_\alpha \overline{\S[\phi_\alpha^*,\phi_\alpha]}-\frac{\bar{g}}{2}\sum_{\alpha\beta}\int_{\bx,x_0,x_0'}|\phi_\alpha(x_0)|^2|\phi_\beta(x_0')|^2,
\label{Srep}
\end{equation}
 the first term corresponds to $n$ identical copies of the disorder- averaged action with $\alpha$ the replica indices.  Note that also for the disorder-averaged 
coefficients $\bar{\gamma}_1=-\partial \bar{r}/\partial x$ and $\bar{\gamma}_2=-\frac 12\partial^2 \bar{r}/(\partial x)^2$. The replica-mixing disorder vertex arises from the quadratic 
order of the cumulant expansion and is given by the variance of the mass distribution, 
\begin{equation}
\bar{g}=(\Delta r)^2=\overline{(r_i-\bar{r})^2}.
\end{equation}  
 
As is evident, it is non-diagonal in imaginary time as a consequence of the perfect correlation of the disorder along the imaginary  time direction. Since the disorder also couples to the temporal 
gradient terms, the replica theory contains disorder vertices with additional time derivatives.\cite{Mukhopadhyay+96,Weichman+08} These terms turn out to be irrelevant and do not 
change the universality of the transitions between the MI and the BG and hence are not considered any further.

As a consequence of the underlying strong-coupling expansion, the variance $(\Delta r)^2$ of the mass distribution is not simply given by the variance 
of the random-site disorder in the BH model.\cite{Kruger+09} This is easily seen by expanding $r_i=r(x-\epsilon_i/U)$ for small $\epsilon_i/U$ yielding 
$(\Delta r)^2=\gamma_1^2 \overline{\epsilon^2}/U^2+\gamma_2^2(\overline{\epsilon^4}-\overline{\epsilon^2}^2)/U^4+\ldots$ with $\overline{\epsilon^n}$ 
the $n$-th moment of $P(\epsilon)=P(-\epsilon)$. Hence, the effective disorder $\bar{g}$ is a continuous function of the chemical potential entering via the 
coefficients $\gamma_{1,2}$ and is strongly suppressed at the tips of the MI lobes where $\gamma_1=0$. Therefore, the inclusion of higher moments of the
distribution $P(\epsilon)$ is essential not only to account for the boundedness of the distribution but also to capture the coupling to disorder at commensurate fillings. 
In order to retain the moments to \emph{infinite} order, without loss of generality, we use the discrete distribution 
\begin{equation}
P(\epsilon)=(1-2p)\delta(\epsilon)+p\delta(\epsilon-\Delta) +p\delta(\epsilon+\Delta),
\end{equation} 
corresponding to on-site energies which occur with the same probability $p<1/2$ increased or decreased by $\Delta<U/2$.  The resulting mean $\overline{r}$ and variance
$\overline{g}$ of the induced mass distribution entering the replica theory (\ref{Srep}) are given by 
\begin{eqnarray}
\overline{r} & = & \overline{r(x-\epsilon_i/U)}\nonumber\\
& = & (1-2p)r(x)+p[r(x-\delta)+r(x-\delta)],\\
\overline{g} & = & \overline{[r(x-\epsilon_i/U)-\overline{r})]^2}\nonumber\\
& = & (1-2p)r^2(x)+p[r^2(x-\delta)+r^2(x+\delta)]-\overline{r}^2,
\end{eqnarray}
and contain the moments of $P(\epsilon)$ to infinite order. The MF phase boundary between the $m=1$ MI and the SF obtained by the condition $\overline{r}=0$ is shown in Fig. 1(a). 
Whereas at incommensurate fillings the interaction vertex $\bar{h}$ is irrelevant for $D>2$, the disorder vertex is always relevant in $D<4$ rendering the MF theory meaningless. In order to determine the stability region of the MI and subsequent phase transitions, we employ the RG.

\section{Renormalization Group Analysis}
\label{sec.RG}

We proceed with the RG analysis of the effective replica field theory (\ref{Srep}). From the RG flow of the mean $\bar{r}$ and the variance $\bar{g}$ of the
disorder-induced mass distribution we can subsequently determine the scale dependence of the relative variance $R_r=\bar{g}/\bar{r}^2$ to analyze the instability of the MI towards the 
formation of a BG.

\subsection{Renormalization Group Equations}

After successively eliminating modes of highest energy corresponding to momenta from the infinitesimal shell $e^{-\ud \ell}\le|\bk|\le1$ and rescaling of momenta ($\bk\to\bk e^{\ud \ell}$), 
frequencies ($\omega\to\omega e^{z\ud \ell}$), and fields  ($\phi_\alpha\to\phi_\alpha^{-\lambda\ud \ell}$), we obtain to one-loop order\cite{Kruger+09} 
\begin{subequations}
\begin{eqnarray}
\frac{\ud \bar{r}}{\ud \ell} & = & 2\bar{r}+2I_1\tilde{h}-I_0 \tilde{g}, \\
\frac{\ud \bar{\gamma}_1}{\ud \ell} & = & (2-z)\bar{\gamma}_1+I_0^2 \bar{\gamma}_1 \tilde{g},\\
\frac{\ud \bar{\gamma}_2}{\ud \ell} & = & (2-2z)\bar{\gamma}_2+I_0^2(I_0 \bar{\gamma}_1^2+\bar{\gamma}_2)\tilde{g},\\
\frac{\ud \tilde{h}}{\ud \ell} & = & (4-D-z)\tilde{h}-(J_0+4J_1)\tilde{h}^2+6I_0^2\tilde{g}\tilde{h},\\
\frac{\ud \tilde{g}}{\ud \ell} & = & (4-D)\tilde{g}+4I_0^2\tilde{g}^2-4J_1\tilde{g}\tilde{h},
\end{eqnarray}
\label{RGeqs}
\end{subequations}
where the scaling dimension $\lambda$ of the fields has been determined such that the coefficient of the spatial gradient term remains constant. Further,
we have absorbed a numerical factor arising from the shell integration by redefining $\tilde{h}=2S_D/(2\pi)^D\overline{h}$ and $\tilde{g}=S_D/(2\pi)^D\overline{g}$
with $S_D$ the surface of the $D$-dimensional unit sphere. The frequency integrals are defined as $I_0=\C(0)=1/(1+\bar{r})$, 
$I_1=\int_\omega\C(\omega)=[\bar{\gamma}_1^2+4\bar{\gamma}_2(1+\bar{r})]^{-1/2}$, $J_0=\int_\omega\C^2(\omega)=I_0I_1/2$, 
and $J_1=\int_\omega|\C(\omega)|^2=2\bar{\gamma}_2I_1^3$ with $\int_\omega=\int_{-\infty}^\infty\frac{\ud\omega}{2\pi}$ and 
$\C(\omega)=1/(1+\bar{r}-i\bar{\gamma}_1\omega+\bar{\gamma}_2\omega^2)$ the on-shell propagator.

\subsection{New Variables}

From the RG equations, it is evident that for $D<4$, the variance $\tilde{g}$ increases with the scale which suggests that the RG flow is always towards strong disorder. 
This interpretation is inconsistent with perturbative arguments which show that the disordered BH model exhibits stable MI phases in the regime of weak disorder,\cite{Fisher+89}
and does not take into account that in the insulating phases the mean $\bar{r}$ also diverges under the RG. Therefore the question of whether disorder is strong or not can only be answered 
by analyzing the relative variance $R_r\sim \tilde{g}/\bar{r}^2$. In the stable MI phase, disorder is irrelevant and we expect that $R_r(\ell)\rightarrow 0$.

Realizing that $\bar{r}$ diverges for both insulating phases, we implement the approach of Aharony and Harris\cite{Harris+96} by focusing on the relative variance 
$R_r$ of the disorder-induced mass distribution. As we argued in Sec. \ref{sec.selfaveraging}, $R_r$ serves as the order parameter for the MI/BG transition characterized by a breakdown 
of self-averaging. 

Putting this into practice in the RG equations requires a new set of variables. Instead of the mass $\bar{r}$, we use $I_0= 1/(1+\bar{r})$ which corresponds to the on-shell propagator 
in the static limit, $\C(0)$, and asymptotically becomes the inverse mass in the insulating phases where $\bar{r}(\ell)\to\infty$.  In addition, we work with 
$\Gamma_i=I_0\gamma_i$ ($i=1,2$) for the temporal gradient terms and $H=I_0^2\tilde{h}$, $G=I_0^2\tilde{g}$ for the interaction and disorder vertices, respectively.  
In the following, we identify the MI phase by the conditions that $\bar{r}(\ell)\to\infty$ but $R_r(\ell)\to 0$.  In the BG, $R_r(\ell)\ne 0$. In the large mass regime, $G$ corresponds precisely to 
the relative variance $R_r$ of the mass distribution.

\section{Results}
\label{sec.results}

\subsection{Numerical Integration}

We start with a numerical integration of the full set of RG equations (\ref{RGeqs}) in terms of the variables introduced above to see if we can indeed identify the MI/BG transition.
Since $\bar{\gamma}_1\neq 0$ away from the tips of the Mott lobes, we use a dynamical exponent $z=2$.  Note that for small $\bar{r}$ close to the MI-SF transition in the clean system, 
the spatial gradient term $\gamma_x=1$  serves as a cut-off in $\C(\omega)$ and the corresponding frequency integrals.  

Indeed we find a regime where the system flows 
towards a stable MI fixed point $P_\textrm{MI}$ with $I_0=0$ and $G=0$. Therefore, in the MI phase the mean $\bar{r}$ of the mass distribution diverges fast enough that the 
relative variance $R_r$ of the distribution renormalizes to zero. As expected, $\Gamma_{1,2}$ and $H$ renormalize to zero in the MI. 

Using the analytic expressions for the bare coefficients as a function of the microscopic parameters, we obtain the MI phases in the presence of disorder. Fig. 1(a) shows 
the resulting $m=1$ MI lobe in $D=3$ as a function of $x=\mu/U$ and $y=2Dt/U$ for $p=0.1$ and $\delta=0.2$.  Sufficiently close to the MI state, $G$ diverges 
at a certain scale $\ell^*$ whereas $I_0(\ell^*)$ is practically zero. This clearly identifies the adjacent phase as an insulating BG! Note that the mass distribution 
spreads faster than it shifts to infinity. Hence it is the tail of the distribution which destabilizes the MI, indicative of rare-region physics.

\begin{figure}[t]
\includegraphics[width=\linewidth]{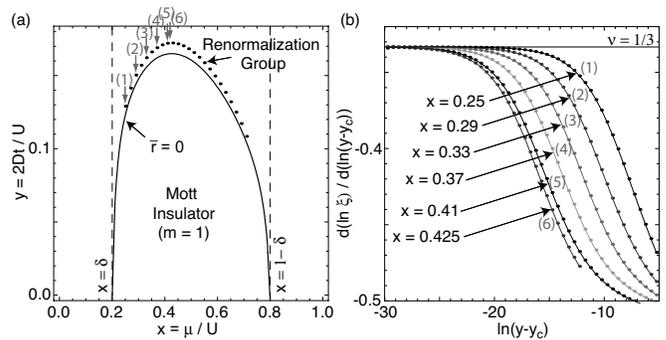}
\caption{(a) Phase boundary between the MI ($m=1$) and the BG obtained from numerical integration of the RG equations. The disorder corresponds to 
site energies increased or decreased by $\delta=\Delta/U=0.2$ with probability $p=0.1$. The solid line shows the unphysical mean-field phase 
boundary ($\bar{r}=0$) between the MI and SF. (b) Extraction of the correlation length exponent for different values $x$ of the chemical potential as indicated in (a).}
\label{fig.1}
\end{figure}

In the following, we determine the phase boundary $y_c(x)$ by bisection  and follow the divergence of the correlation length $\xi=e^{\ell^*}$ as $y\to y_c^+$. Note that the correlation length 
defined here is determined by the scale $\ell^*$ where $R_r$ diverges and self-averaging breaks down. The correlation length $\xi$ therefore corresponds to the 
distance of rare regions in the BG which cause the breakdown of self-averaging (see Sec. \ref{sec.selfaveraging}). Sufficiently close to the MI, we expect a power-law divergence, 
$\xi\sim(y-y_c)^{-\nu}$, defining the correlation-length exponent $\nu$. To extract $\nu$, in Fig. 1(b) we plot $\partial(\ln\xi)/\partial(\ln(y-y_c))$ as a function 
of $\ln(y-y_c)$ for different values of $x$. For $y\to y_c^+$, the functions indeed converge to a constant value corresponding to $\nu=1/3$. 

Interestingly, on approaching commensurate boson fillings,  the length scale upon which universal behavior is observed increases and possibly even diverges. Whereas an increase 
might be simply a quantitative effect originating from the suppression of the bare disorder vertex by commensuration, a divergence would indicate a different universality class at the tips 
of the MI lobes.

\subsection{Universality of the MI/BG Transition at Incommensurate Filling}

To better understand the nature of the transitions  we proceed with an analytical investigation of the \emph{incommensurate} case ($z=2$). Since for $D>2$ the interaction 
vertex $\bar{h}$ is irrelevant at the MI-SF transition of the clean system, in the vicinity of the fixed point $P_\textrm{MI-SF}$ ($G=0$, $I_0=1$) we can simplify the RG equations 
by setting $\tilde{h}=0$. The resulting RG equations
for $I_0$ and $G$, 
\begin{subequations}
\begin{eqnarray}
\frac{\ud I_0}{\ud \ell}  & = &   (G-2)I_0 +2I_0^2,\\
\frac{\ud G}{\ud \ell} &  = &  (4I_0-D)G+6G^2,
\end{eqnarray}
\label{RGeqsIC}
\end{subequations}
are decoupled from $\Gamma_{1,2}$. Our analysis of these equations is
strictly in the perturbative regime where $G(\ell)\le 1$.  The
divergence of $I_0$ which obtains for $G(\ell)>2$ is spurious and
outside the range of validity of the perturbative RG equations.  In addition to $P_\textrm{MI-SF}$ and the stable MI fixed point $P_\textrm{MI}(I_0=0,G=0)$, the RG equations (\ref{RGeqsIC}) exhibit an 
additional fixed point  $P_\textrm{MI-BG}(0,D/6)$, which will turn out to be critical fixed point determining the universality of the MI-BG transition at incommensurate 
fillings. From linearization around $P_\textrm{MI-BG}$ we find $\ud H/\ud \ell=(D/3-2)H$ indicating that the interaction vertex remains irrelevant. 

For $G=0$ we obtain a trivial flow equation for $I_0$ capturing the MI-SF of the clean system. For $I_0<1$, the flow is towards $P_\textrm{MI}$, while for $I_0>1$ ($\bar{r}<0$), 
$I_0(\ell)\to\infty$ signaling the instability toward the formation of a SF.  

For $D<4$, $P_\textrm{MI-SF}$ is unstable for small  disorder, $G>0$. However, instead of a runaway flow, we find a separatrix $G_\textrm{s}(I_0)$ which connects 
$P_\textrm{MI-SF}$ and $P_\textrm{MI-BG}$ as shown in the case $D=3$ in Fig. 2(a). Linearizing around the fixed-points, we find that asymptotically 
$G_\textrm{s}(I_0)\simeq(D-2)(1-I_0)$ in the vicinity of $P_\textrm{MI-SF}$ and $G_\textrm{s}(I_0)\simeq D/6-4I_0/(12-D)$ close to $P_\textrm{MI-BG}$, respectively. 
 
For initial values $I_0(0)<1$ and $G(0)<G_\textrm{s}(I_0(0))$, the system flows towards the MI fixed point, whereas for $G(0)>G_\textrm{s}(I_0(0))$, $G(\ell)$
will eventually diverge.  For initial values very close to the MI phase, the trajectory $(I_0(\ell),G(\ell))$ will track the separatrix and consequently, sufficiently close to the transition,  
the divergence of $G(\ell)$ is controlled by the fixed point $P_\textrm{MI-BG}$. Linearizing around this point, we find $G(\ell)=(\Delta G)e^{D\ell}$ with $\Delta G=G_0-G_\textrm{s}$ 
the infinitesimal distance to the separatrix. From the condition $G(\ell^*)\simeq 1$ we obtain the correlation length $\xi\sim (\Delta G)^{-1/D}$, implying a correlation length exponent 
of $\nu=1/D$, in perfect agreement with the value extracted from numerical integration of the full set of RG equations for $D=3$ (see Fig. 1(b)) and a previous 
real-space RG and scaling analysis.\cite{Pazmandi+98,Pazmandi+97}

\begin{figure}[t]
\includegraphics[width=\linewidth]{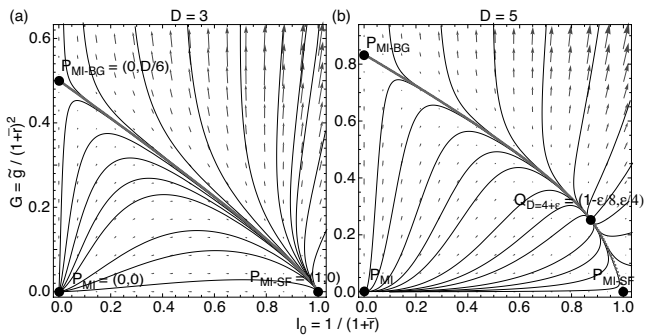}
\caption{RG flow for (a) $D=3$ and (b) $D=5$ for incommensurate boson fillings as a function of $I_0$ and $G$ corresponding (in the limit of large mass $\bar{r}$) to the 
inverse mean and the relative variance of the mass distribution, respectively.}
\label{fig.2}
\end{figure}

Since $G\ne 0$ at the fixed point and in the BG, the fluctuations are no longer\cite{Harris+96} governed by the central-limit theorem.  This implies that self-averaging 
breaks down.  Consequently, the bound $\nu\ge 2/D$ [Ref. \onlinecite{Chayes+86}] no longer applies. That the BG is mediated by rare localized regions is the efficient cause 
of this effect. In quantum systems, such rare events play a more pronounced\cite{Vojta+03} role than in classical systems because of the perfect correlation of the disorder along the 
imaginary time direction.

It is also the divergence of the relative variance that is responsible for a finite compressibility $\kappa=(\beta U)^{-1}\partial^2\overline{\ln Z}/\partial \bar r^2$ in the BG. 
Using the asymptotic form of $\S_\textrm{eff}$ in which only the $\bar r$ and $\bar g$ terms are retained, we obtain to leading order $\kappa\approx 1/\bar r^2(1+4\bar g/\bar r^2)$ which clearly vanishes in the MI. However, in the BG 
sufficiently close to the transition, $\kappa\equiv \kappa(l^*)\sim\xi^{D-4}\sim(y-y_c)^\gamma$ with $\gamma=4/D-1$.  

For completeness, we briefly discuss the case $D>4$ in which a small amount of disorder is irrelevant at the MI-SF transition. The inversion of the RG flow along the $G$-direction
is a consequence of the presence of an additional unstable fixed point $Q(1-\epsilon/8,\epsilon/4)$ in $D=4+\epsilon$ located on the separatrix (see Fig. 2(b)). Therefore, 
in $D>4$ the disorder has to exceed a critical value in order to
induce a BG phase in contrast to recent claims.\cite{Pollet+09}

\subsection{Commensurate Filling}

We now focus on \emph{commensurate} boson fillings to understand the nature of the transition at the tips of the MI lobes where $\bar{\gamma}_1=0$ and $z=1$. The smaller
dynamical exponent makes the interaction vertex $H=I_0^2\tilde{h}$ more relevant and in fact marginal at the MI-SF transition of the clean system in $D=3$. However, for 
$D=4-\epsilon$, $H$ remains irrelevant at $P_\textrm{MI-SF}$ suggesting that the critical behavior is controlled by the same separatrix as for incommensurate fillings. 
Interestingly, this is not the case. A numerical integration of the RG equations shows that in the disordered phase close to the MI, $H(l)$ diverges 
simultaneously with $G(l)$. This behavior is easily understood upon linearizing around $P_\textrm{MI-BG}$ yielding $\ud H/\ud l=(D/3-1)H$ which demonstrates that 
$H$ becomes relevant for $D>3$. Therefore, for any finite value $H(0)>0$, the separatrix obtained by projection into the $I_0$-$G$ plane does not terminate 
in $P_\textrm{MI-BG}$ but diverges as $G_\textrm{s}\sim I_0^{-D/(12-D)}$. Consequently, the MI-BG transition at commensurate fillings is not controlled by $P_\textrm{MI-BG}$
but by a different fixed point which is not accessible in the present calculation. However, from the divergence along the seperatrix, $G_\textrm{s}\sim e^{\frac{D}{6} l}$ we estimate $\nu=6/D$ 
which is significantly larger than the incommensurate value and satisfies the bound $\nu\ge 2/D$.\cite{Chayes+86}

\section{Final Remarks}
\label{sec.final}

To summarize, we have analyzed the instabilities of the MI state of the weakly disordered BH model within an RG analysis of the disorder-averaged
replica theory valid in the strong-coupling regime. Our analysis shows that the MI always becomes unstable towards a BG regardless of the boson filling, in agreement with recent 
work\cite{Pollet+09} which, based on a mathematical theorem, excludes the possibility of a direct MI/SF transition in the presence of disorder. 

While in both insulating phases the mass coefficient diverges on large scales, rendering SF correlations short ranged, the relative variance of the induced random mass distribution 
renormalizes to zero in the MI and diverges in the BG. This allowed us
to demonstrate that the relative variance serves as the order parameter for the MI/BG transition. We further 
extracted the diverging correlation length of the MI/BG transition from the scale where the relative variance diverges under the RG. Since the finite relative 
variance of an extensive quantity signals a breakdown of the central-limit theorem,\cite{Harris+96} the correlation length corresponds to the separation between rare regions in the BG, causing a breakdown 
of self-averaging in the system. We point out that the correlation length defined here has nothing to do with the SF correlation length which remains
finite over the MI/BG transition.

The reformulation of the RG equations in terms of
new variables, in particular the relative variance, allows us to identify fixed points and to subsequently analyze 
the universality of the MI/BG transition. That new variables, not
apparent in the UV-complete theory, are necessary to access the fixed
point is typical of strong-coupling problems. At incommensurate fillings, we find a separatrix which connects the MI/SF fixed point in the clean system with the critical MI/BG fixed point in the
weakly disordered system. The correlation-length and compressibility exponents characterizing the MI/BG transition are determined as $\nu=1/D$ and $\gamma=4/D-1$, respectively. 
In principle, the power-law divergence of the inverse compressibility could be verified either experimentally or numerically provided that the critical region can be accessed sufficiently.
The violation of the strict lower bound $\nu\ge 2/D$ known as the quantum Harris criterium\cite{Chayes+86} has been argued some time ago\cite{Harris+96,Pazmandi+97} to be a 
key signature for the breakdown of self averaging. Interestingly, our correlation-length exponent $\nu=1/D$ is identical to the exponent estimated by a scaling analysis in the atomic limit
for a box distribution of site energies.\cite{Pazmandi+97} 

Our analysis shows that the MI/BG transition at commensurate fillings is controlled by a different fixed point which is not accessible within the present 
RG approach. The two distinct transitions found here are, in retrospect, not out of the ordinary, since in the clean system, the MI/SF transitions are in the universality 
class of the $(D+1)$-dimensional XY model\cite{Doniach81,Fisher+88} and of mean-field type\cite{Fisher+89} for commensurate and incommensurate fillings, respectively.  
This dichotomy persists in the disordered system as well.

What this analysis indicates is that within a replica theory, instabilities driven by local rare regions, indicative of Griffiths singularities,
are completely accessible if one focuses on the scaling dependence of the relative variance of the mass distribution.  Since the relative
variance of any extensive quantity acquires a non-zero value \cite{Harris+96} once self-averaging breaks down, the method
presented here can in principle be applied to any transition driven by local rare-region physics.

\textbf{Acknowledgment:} We are grateful for stimulating discussions
with E. Fradkin, A.~J. Leggett, and A.~G. Green and we thank the
NSF DMR-0940992, the Samsung Scholarship (S. H.), the DARPA OLE program, and the
Macarthur Professorship endowed by the John D. and Catherine
T. Macarthur Foundation at the University of Illinois (F. K.) for financial support.

\end{document}